\documentclass[aps,prd,showpacs,nofootinbib,twocolumn,superscriptaddress,floatfix]{revtex4}

\usepackage{amsmath,amssymb}%,graphicx}
\usepackage[dvips, pdftex]{graphicx}
\usepackage[tight,footnotesize]{subfigure}
\usepackage[usenames]{color}
\usepackage[dvipsnames]{xcolor}
\usepackage{float}
\usepackage{wrapfig}
\usepackage{hyperref}
\usepackage{mathtools}

\DeclareGraphicsExtensions{.jpg,.mps,.pdf,.png,.eps} 

\renewcommand{\ij}{{}_{ij}} 
\renewcommand{\IJ}{{}^{ij}}

\newcommand{\M}{\mathcal{M}}

\newcommand{\gm}{\gamma}

\newcommand{\lp}{\left(}
\newcommand{\rp}{\right)}

\newcommand{\tgm}{{\tilde\gm}}

\newcommand{\tA}{{\tilde  A}}

\newcommand{\be}{\begin{equation}}
\newcommand{\ee}{\end{equation}}
\newcommand{\bea}{\begin{eqnarray}}
\newcommand{\eea}{\end{eqnarray}}

\newcommand{\rr}{\mathrm}

\newcommand{\mpl}{m_\rr{pl}}
\newcommand{\Mpl}{M_\rr{pl}}

\begin{document}
\title{
Beginning inflation in conformally curved spacetimes
}

\author{Cristian Joana}
\email{cristian.joana@itp.ac.cn}
\affiliation{CAS Key Laboratory of Theoretical Physics, Institute of Theoretical Physics, Chinese Academy of Sciences, Beijing 100190, China }

\affiliation{Cosmology, Universe and Relativity at Louvain (CURL),
	Institut de Recherche en Mathematique et Physique (IRMP),
	University of Louvain,
	2 Chemin du Cyclotron,
	1348 Louvain-la-Neuve,
	Belgium}

\pacs{98.80.Cq, 98.70.Vc}
	
\date{\today}

\begin{abstract}

We investigate the initiation of cosmic inflation, in full numerical relativity, from pre-inflationary scenarios with large tensor and vector fluctuations in the metric. These settings are characterized by having large values in the Weyl curvature tensor. In the matter sector, we consider a single scalar field with inhomogeneous field velocities, corresponding to a kination period. In the context of large-field inflation, it is shown that the onset of inflation continues to be robust to this type of initial conditions, and that during inflation the Universe successfully homogenizes and flattens any type of curvature, leading to a Friedmann-Lemaître-Robertson-Walker  Universe after just a few tens of efolds of accelerating expansion. 
\end{abstract}

\maketitle

\section{Introduction}\label{sec:intro}

Cosmic inflation \cite{STAROBINSKY198099, PhysRevD.23.347, 10.1093/mnras/195.3.467, LINDE1982389} was originally proposed to explain the observed  large scale homogeneity and flatness of today's Universe.
This theory postulates an early phase of rapid and accelerated expansion of the Universe. During inflation, quantum fluctuations became red-shifted, exiting the Hubble volume at the time, what yields a nearly scale-invariant power spectrum of cosmological perturbations that matches the observations of the cosmic microwave background (CMB) \cite{Akrami:2018odb,Ade:2015lrj}. At later times, these perturbations would constitute the seeds for structure formation and, under certain theoretical modelling, it also allows for the 
the generation of primordial black holes as promising dark matter candidates \cite{LISACosmologyWorkingGroup:2023njw}. 

Returning to the original purpose of the theory, cosmic inflation is able to explain the Universe homogeneity across $10^5$ Hubble volumes at the time of the last scattering. 
However, such an achievement would arguably lose part of its appealing 
if large amounts of homogeneity and/or anisotropy would be required for inflation to begin with, and therefore be fine-tuned to overly specific, or non-generic initial conditions. Since its conception, the generality of the initial conditions required to trigger inflation has often been a topic of controversy~\cite{Penrose:1988mg,Goldwirth:1989pr, Goldwirth:1990pm,Laguna:1991zs,KurkiSuonio:1993fg,Deruelle:1994pa}, and more recently in Refs.~\cite{Martin:2013nzq,Ijjas_2013,Guth_2014,Easther:2014zga,Ijjas_2016,Finn:2018krt,Chowdhury:2019otk,%
Tenkanen:2020cvw,Garfinkle:2023vzf,Ijjas:2024oqn,Martin:2024qnn}.

The issue of initial conditions for inflation has been studied extensively using analytical and numerical approaches (see Ref.~\cite{Brandenberger:2016uzh,Joana2020} for nice reviews). Several works using full numerical relativity simulations have also been done testing for various types of pre-inflationary settings in the non-perturbative regime~\cite{Ijjas:2022qsv}. The robustness of small and large field inflation models was studied in Ref.~\cite{Clough:2017ixw}; the differences between concave and convex potentials were investigated in Ref.~\cite{Aurrekoetxea_2020}. Furthermore, scenarios with large scalar field gradients \cite{East:2015ggf}, inhomogeneous kination \cite{Joana2020,Joana2022}, with large extrinsic curvature perturbations \cite{Clough_2018} and with non-vanishing momentum  have also been investigated~\cite{Corman:2022alv,Elley:2024alx}. 
The working case of the non-minimal Higgs inflation model with a curvaton, or auxiliary field, during inhomogeneous pre-inflation and preheating was considered in Ref.~\cite{Joana2022}.

From all the above-mentioned works, we have gathered many insights into the highly non-linear dynamics of what could have been the pre-inflationary era. On the question of whether inflation requires specific (homogeneous) initial conditions, it has been consistently found that
while small-field inflation models (i.e. operating at sub-Planckian field values) could \textit{suffer} 
to begin from homogeneous~\cite{Chowdhury:2019otk} and inhomogeneous configurations, these works have also shown that large-field inflation (i.e. operating at super-Planckian field values) remain very robust to large inhomogeneities of any kind.
% and the reason for this will be reinforced later in this work. 

However, in a series of papers comparing cosmic inflation with other alternative smoothing mechanisms ~\cite{Cook:2020oaj,Ijjas:2023dnb,Garfinkle:2023vzf,Ijjas:2024oqn}, the authors claimed that inflation could not explain the overall flatness and homogeneity of our Universe, neither for the case of large-field models of inflation. In particular, their claim is that inflation can only  smooth the inhomogeneous Universe if initial conformal flatness is already assumed, 
what would violate Penrose's Weyl Curvature Hypothesis ~\cite{Penrose:1979azm,Penrose:1988mg}. 
The authors also reinforced their conclusions with numerical relativity simulations in \cite{Garfinkle:2023vzf,Ijjas:2024oqn}. Still, their strong claims seem to be in tension with previous works and the later results in Ref.~\cite{Elley:2024alx}.
%\\

In this paper, I present a set of full general relativity simulations investigating the robustness of inflation to begin from inhomogeneous configurations containing large metric fluctuations. Thus, testing a new set of initial conditions where the initial metric is conformally curved. As proposed in Ref.~\cite{Ijjas:2023bhh}, we use the Weyl curvature as a diagnostic to assess the degree of metric inhomogeneities and probe whether a prolonged period of inflation is still a robust outcome, and that it is able to smooth out these curvatures. We find that our conclusions significantly differ from the recent studies \cite{Garfinkle:2023vzf,Ijjas:2024oqn}, and which we will discuss throughout this work.

The organization of the manuscript is as follows: %
Section~\ref{sec:NumStrategy} explains the numerical strategy of the simulations, the theoretical formalism, methodology for constructing the initial data and other technical details. The results of our simulations are explained in Section~\ref{sec:Results}, followed by the discussion in Section~\ref{sec:Discussion}.
Additional information on the notation, code performance, initial data sets and supplementary figures are available in the appendixes.  
In this work I assume a "mostly plus" metric signature $(-+++)$, Greek indices correspond to spatio-temporal dimensions going from 0 to 3, while Latin indices correspond uniquely to spatial indices running from 1 to 3. All quantities are express in Planck units with gravitational constant, speed of light and Planck's mass equal to unity, i.e. $G=c=m_{\rm pl} = 1$, respectively, and $M_{\rm pl} \equiv \frac {m_{\rm pl}}{\sqrt{8\pi}}$ denotes the reduced Planck's mass. \\

\section{Numerical strategy} \label{sec:NumStrategy}

We assume a pre-inflationary era starting nearly after sub-Planckian energies with an average Hubble parameter close to unity. 
If the scalar field potential is subdominant, then the specific form of the potential has little effect on the dynamics of the pre-inflationary epoch, however, it will determine its duration as it determines the energy scales at which inflation might begin. For simplicity, and to facilitate the comparison with Refs.~\cite{Garfinkle:2023vzf,Ijjas:2024oqn}, I consider the well studied case of $V(\varphi) = \frac 12 m^2 \varphi^2$ with a mass set to  $m = 10^{-5}  \mpl$.  
However, similar results are expected for other single-field potentials operating at large field values (i.e. $R^2$-inflation, Higgs inflation, etc.).

\subsection{Numerical formalism}

Similarly as in previous works, we use the Baumgarte-Shapiro-Shibata-Nakamura (BSSN) formalism of General Relativity \cite{PhysRevD.52.5428,Baumgarte_1998}. We solve the Hamiltonian and momentum constraint for the construction of the initial data and  the BSSN evolution equations for the subsequent time-integration. In all generality, the metric reads 
\be \label{timeline}
ds^2 = \alpha dt^2 + \chi^{-1} \tilde \gamma_{ij} (dx^i + \beta^i dt)(dx^j + \beta^j dt) ~,
\ee

where $\tilde\gamma_{ij}$ is the conformal 3-metric, $\chi$ a scalar conformal factor, and $\alpha$, $\beta^i$ are the lapse and shift, receptively, which define the perpendicular vector to the 3-dimensional hypersurface, $n^\mu = (1/\alpha, - \beta_i/\alpha)$. 
In BSSN, the extrinsic curvature $K_{ij}$ is typically rewritten in terms of its conformal trace $K\equiv \chi^{-1}\tilde \gamma^{ij} K_{ij}$ and traceless components 
$ \tilde A_{ij} = \chi^{-1} \left( K_{ij} - \tilde \gamma_{ij} K \right)$.  In this formalism, $K$ relates to the expansion rate of the Universe so that the local Hubble rate can be defined as $H \equiv - \frac 13 K$. 

We can now write a generalised (inhomogeneously localized) form of the Friedmann equations, which can be derived from the Hamiltonian constraint, which reads 
\be   \label{eq:friedman}
H^2 = \frac 1 {3\Mpl^2} \left( \rho_\varphi  + \rho_R + \rho_{\rm shear} \right) ~,
\ee
where the scalar field, curvature and shear energy densities are respectively defined as 
\begin{align} \label{eq:densities}
&\rho_{\varphi} \equiv n^\mu n^\nu T_{\mu\nu} = \frac 12 \dot\varphi^2 + \frac 12 (\partial_i\varphi)^2 + V(\varphi) ~,
\\
&\rho_{R} \equiv  - \frac{\Mpl^2}{2} ~ ^{(3)}R  ~,
\\
&\rho_{\rm shear} \equiv  \frac{\Mpl^2}{2} \tilde A_{ij} \tilde A^{ij} ~.
\end{align}
Here the dotted variables ($\dot \ $) denote the derivatives with respect to the cosmic time, and $ ^{(3)}R$ is the Ricci scalar of the 3-metric,  i.e. $\gamma_{ij} = \chi^{-1}\, \tilde\gamma_{ij}$. The shear energy density $\rho_{\rm shear}$ contains both tensor and vector modes in the gravitational sector. In the matter sector, $T_{\mu\nu}$ is the energy-momentum tensor, and we can also define the equation of state $\omega \equiv \frac {p_\varphi}{\rho_\varphi}$, where the pressure is expressed in terms of the trace of the stress tensor, $ p_\varphi = \frac 13 \gamma\IJ S_{ij} $ with  $ S\ij \equiv  \gamma_{i\mu} \gamma_{j\nu} T^{\mu\nu}$.
\\

As proposed by Ref.~\cite{Ijjas:2024oqn}, we use the Weyl curvature $C$ and Chern-Pontryagin invariant $P$ as a measure of the conformal curvature. These are scalar quantities respectively defined by 
\be
C = {\cal C}_{\alpha\beta\delta\lambda} {\cal C}^{\alpha\beta\delta\lambda} \quad 
~, \ 
P = {\cal C}_{\alpha\beta\delta\lambda} ~ ^* {\cal C}^{\alpha\beta\delta\lambda}  \quad 
 ~,
\ee
where ${\cal C}^{\alpha\beta\delta\lambda}$ is the conformal Weyl tensor and $^* {\cal C}^{\alpha\beta\delta\lambda}$ its dual ~\cite{1984ucp..book.....W}. The successful inflationary period should be able to smooth these quantities below $ \sim {\cal O}(10^{-10}) H^2$ by the end of inflation, in order to explain the CMB observations.

\begin{figure}[t!]
	\hspace*{-5mm}    
    \includegraphics[width=8.5cm]{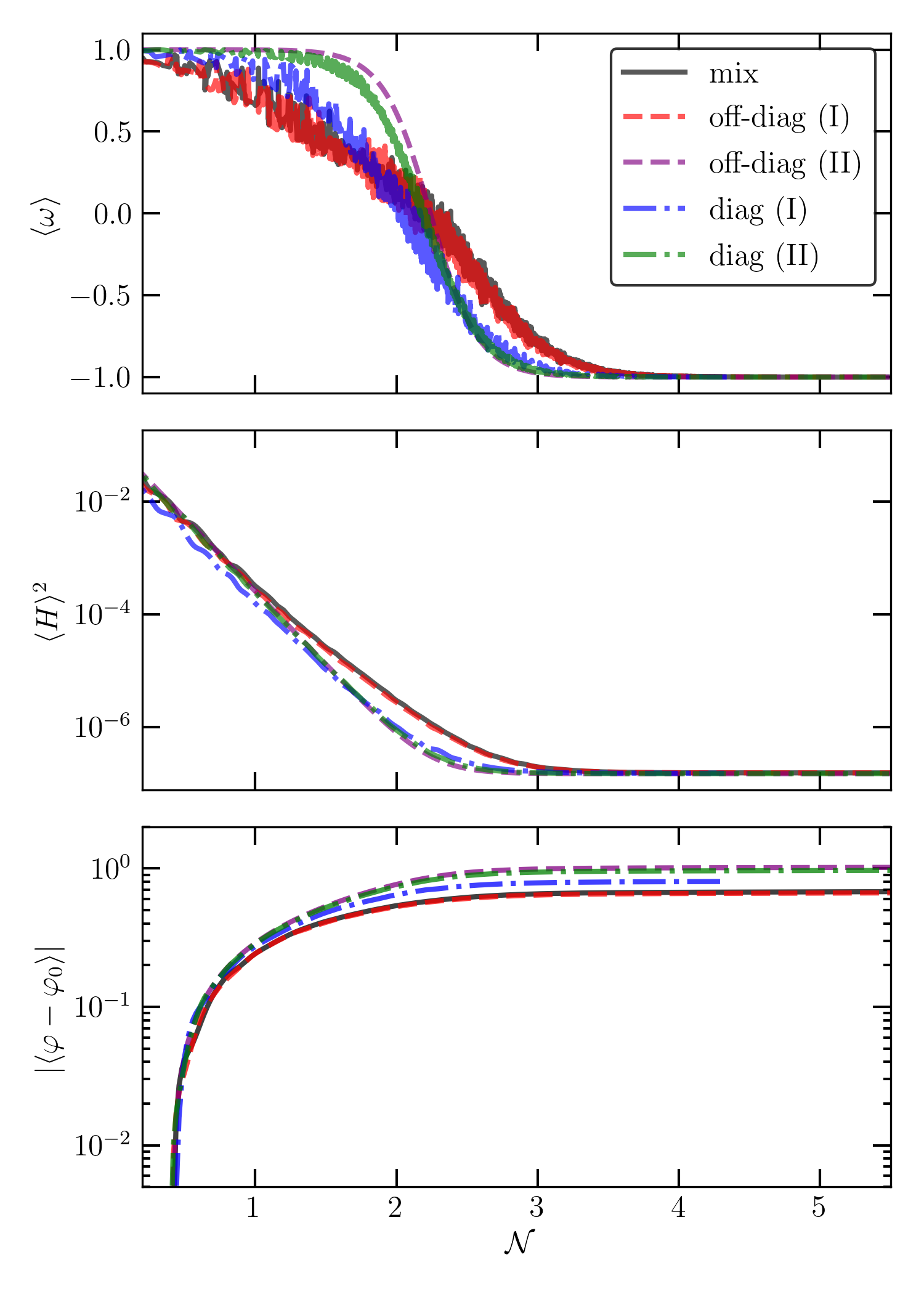} 
    \caption{Evolution of the matter equation-of-sate (top panel), mean Hubble expansion rate (middle panel) and scalar-field excursion with respect its initial value (bottom panel).  }
    \label{fig:sfields}
\end{figure}

\subsection{Construction of initial data}

The initial data is then constructed starting by imposing inhomogeneous configurations for the conformal metric, which is chosen to be $\tilde \gamma\ij (t_{\rm ini}) = \frac 1{{\rm det}(\M\ij)} \M\ij$, with
\be 
\begin{split}
 \M_{11} &= 1 + a_0 + a_1 \Theta(\vec x, a_2, a_3)~,    \quad 
     \M_{12} = d_1 \Theta(\vec x, d_2, d_3)   ~,\\
 \M_{22} &= 1 + b_0 + b_1 \Theta(\vec x, b_2, b_3) ~, \quad \
     \M_{13} = f_1 \Theta(\vec x, f_2, f_3)   ~, \\
 \M_{33} &= 3 + c_0 - M_{11} -M_{22}  ~, \qquad   
     \M_{23} = g_1 \Theta(\vec x, g_2, g_3)   ~,  
\end{split}
\ee
where $ \M_{ji} = \M_{ij}$, and 
\be
\begin{split}
\Theta(\vec x, \lambda, \theta) = & \frac 13 \sum_{i=1,2,3} \cos \left( \frac {2\pi}L \lambda x_i + \theta \right) ~,
\end{split}
\ee
with $a_n$, $b_n$, $c_n$, $d_n$, $f_n$,  $g_n$, for any $n$, are parametric constants. 
The conformal factor $\chi$ is initially set to one everywhere. These choices sets the initial 3-metric, and in turn, also its associated Ricci scalar $^{(3)}R$. In addition, we initially impose (momentously) a vanishing $\tilde A\ij =0$ and a homogeneous value for $K$ (e.g. initial expansion rate). Thereafter, we can solve the constraint equations to get a valid configuration for the scalar field variables. For simplicity, we choose an initial inhomogeneous configuration of kination with 
\be 
\dot \varphi (\vec x, t_{\rm ini}) =  - \sqrt{\Mpl^2 ~ ^{(3)}R   + \frac 23 \Mpl^2  K^2 - 2V(\varphi)}
\ee
while keeping an homogeneous value for the scalar field at $\varphi(\vec x, t_{\rm ini}) = 20 ~\mpl$. These initial configurations are appropriate for our study, as we want to investigate the role of largely dominated tensor and vector modes in the metric. We note that $\tilde A\ij$ is only vanishing at the initial time, and rapidly grows in the subsequent time-steps sourced by to the metric fluctuations. In the matter sector, kination energy tends to decay as 
$\frac 12 \dot\varphi^2 \propto a^{-6}$,
partly generating scalar-field gradients, and yielding to a system with deeply inhomogeneous, sub-Hubble dynamics where 
$C,\ P,\ \rho_{\rm shear} \gtrsim H^2$  and ${\frac 12 (\partial_i \varphi)^2, \frac12 \dot\varphi^2 \gtrsim H^2}$.

\subsection{Numerical details}

The simulations have been done using a grid composed by $(160)^3$ to $(280)^3$ cells with an initial grid-size that is around three times the Hubble radius $L\approx 3H^{-1}$, I also tested increasing the grid side to $L\approx 10H^{-1}$, allowing for  super-Hubble initial metric fluctuations without significant changes. The topology is of a 3-dimensional torus with periodic boundary conditions in all dimensions. %
The evolution of the dynamical system is computed by numerical integration of the BSSN equations \cite{PhysRevD.52.5428,Baumgarte_1998,10.1143/PTPS.90.1} in 3+1 dimensions, implemented in the GRChombo code \cite{Clough_2015,Andrade:2021rbd}. A more detailed explanation of the structure and validation of the code can be found in the supplemental material and in Refs.~\cite{Clough_2015,Andrade:2021rbd}. 
\\

In the following analyses, I denote the mean value of variable $\theta$ at a given time-hypersurface with $\langle ... \rangle$ brackets, which is defined by
\be \label{eqn:Kini}
\langle \theta \rangle \equiv \frac{1}{{\cal V}} \int \theta \, \rr d {\cal V}~,
\ee
where ${\cal V}$ is the spatial volume. Similarly, the root-mean-square (r.m.s.) is given by 
\be
{\rm r.m.s\ of\ } \theta = \sqrt{\langle \theta^2 \rangle}
~, 
\ee

For instance, we can use the previous identities to compute the mean number of efolds $\cal N$ and the scalar curvature $\delta\mathcal{N}$, as it follows
\be
 { \cal N }  = \langle \ln(a) \rangle  
 ~, \quad 
\delta\mathcal{N} =  \sqrt{\langle \ln(a)^2 \rangle - \langle \ln(a) \rangle^2 }  ~,
\ee
with $a = 1/\sqrt{\chi} $ is the local cosmological scale factor as it can be deduced from matching Eq.~(\ref{timeline}) to the Friedmann-Lemaître-Robertson-Walker (FLRW) metric.

\begin{figure*}[t!]
    \includegraphics[width=18.cm]{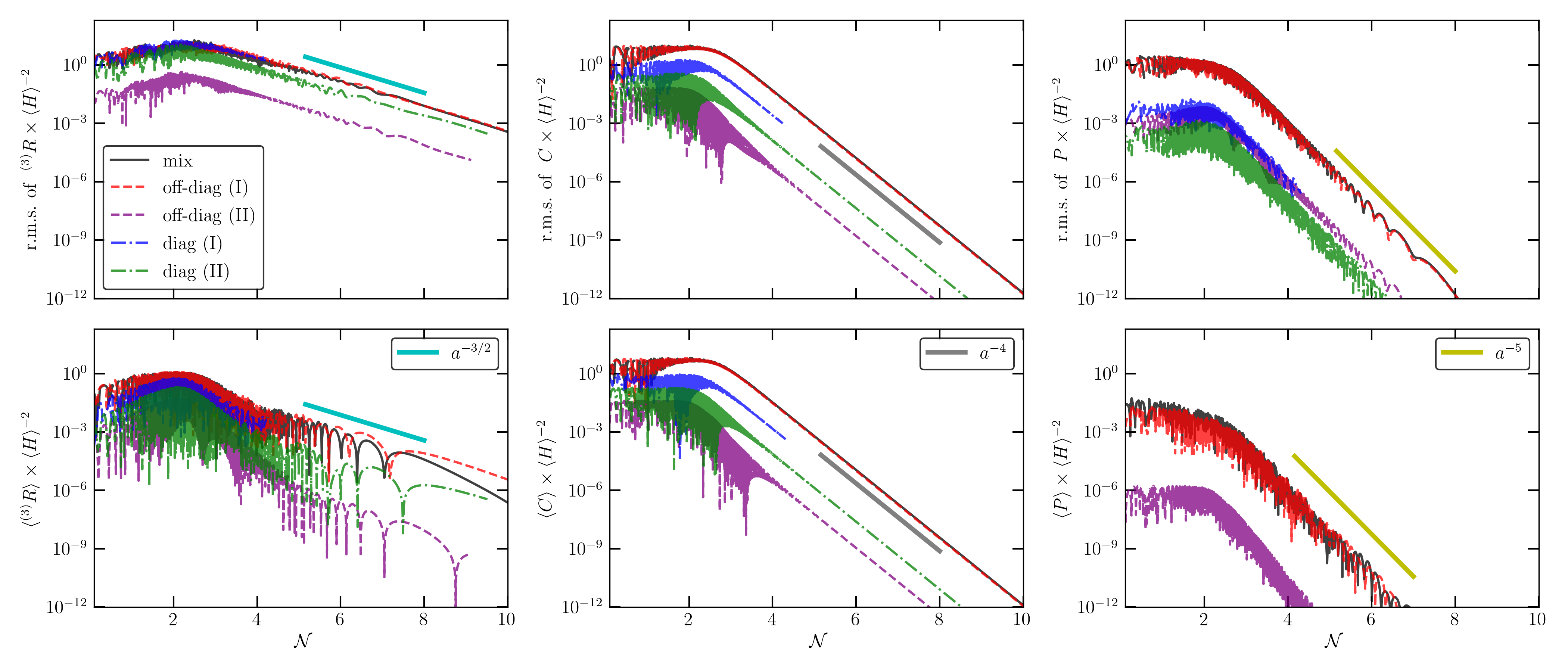} 
    \caption{Top panel: evolution of mean three-dimensional Ricci scalar (left panels), Weyl curvature (middle panels) and Chern-Pontryagin invariant (right panels) for several configurations initially containing vector and tensor (solid lines), vector-only (dashed lines) and tensor-only (dashed-dotted lines) fluctuations. Bottom panel: the same but for the root-mean-squared values of the same quantities. All quantities are normalized by the squared of the mean Hubble rate, and the approximate scaling law after inflation onset is indicated in solid cyan, gray and green-lime lines for each corresponding quantity. }
    \label{fig:evo_curvatures}
\end{figure*}

\begin{figure*}[b!]
    \centering
    \includegraphics[width=17.5cm]{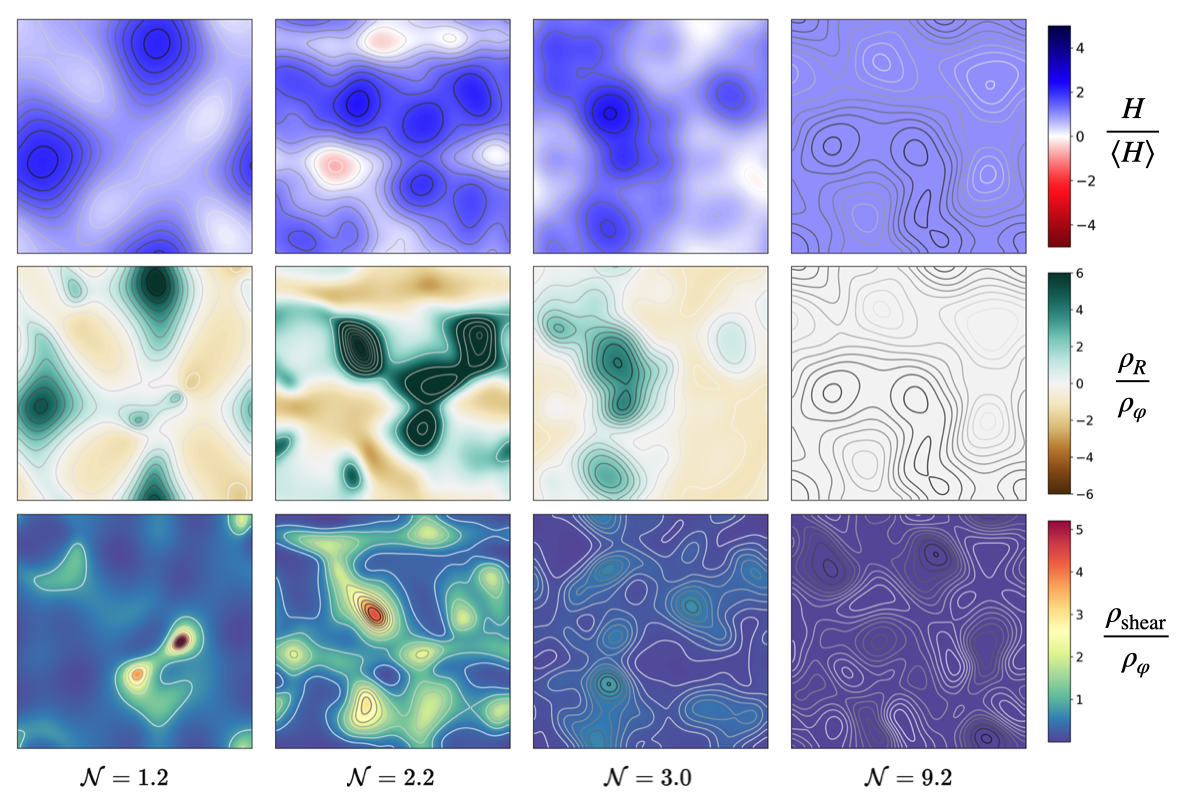} 
    \caption{Two-dimensional snapshots of the lattice configurations of the annotated quantity at the selected efolding time. Top panel: the ratio of local-to-average Hubble rate; middle panel: the ratio of curvature-to-scalar energy density, and bottom panel: the ratio of shear-to-scalar energy density. All figures correspond to a simulation initially containing both tensor and vector gravitational modes represented in a black solid line in figure~\ref{fig:sfields}. Each plot containS ten equally spaced contours between max and min values at the time-hypersurface to help visualising the magnitude of the fluctuations.  }
    \label{fig:pics_density}
\end{figure*}

\section{numerical evolution} \label{sec:Results}

In our simulations, we can distinguish three scenarios depending on the metric fluctuations. The first category is for simulations where in the initial conditions we only introduced \textit{diagonal} metric fluctuations, on the terms $\tilde\gamma_{ii}$, which relate predominantly to gravitational vector (longitudinal) fluctuations. Secondly, we considered simulations starting with only \textit{off-diagonal} metric fluctuations which relate to tensor (transverse) fluctuations. The third category is the \textit{mix} combination of both vector and tensor metric inhomogeneities. %

In Fig.~\ref{fig:sfields}, we show the averaged evolution of the equation-of-state $\omega$, the scalar-field energy density, and the variation of the scalar-field throughout the time evolution. In the homogeneous (kination) limit, the equation of state experiences a fast transition from kination, i.e. $\omega = 1$ to inflation with $\omega \simeq -1$.  
%%%%
%%%
This is still very much the case when considering inhomogeneities where curvatures are sub-dominant, i.e. $\ ^{(3)}R,\ C \lesssim 10^{-1}H^2 $. However, when strong fluctuations are added, especially of the tensorial off-diagonal form, we observe a more gradual transition from kination to inflation reflecting an efficient energy transfer between metric tensor fluctuations and scalar-field gradients. This is in agreement with what was reported in Ref.~\cite{Joana2020}, referred to as ''extrinsic curvature modes". 
%&&&
%%%%%
Another relevant observation is that, despite starting with large scalar-field velocities, these velocities are exponentially suppressed during the pre-inflationary epoch by the Hubble friction. Therefore, we obtain that the largest field variations during this period is of $\Delta \varphi \approx 1~ \mpl$, a value which is further reduced in the presence of strong scalar-field gradients. 

\subsection{Wax and wane of curvature}

%Weyl curvature part 
In our simulations, we set up initial conditions so that they begin with values for the Ricci, Weyl and Chern-Pontryagin scalars of the order of the Hubble parameter squared, i.e. $^{(3)}R, C, P \sim H^2$.  When tensor perturbations are added we easily get large values for $C$ and $P$. However, when only vector fluctuations are included, curvature values are relatively smaller, particularly for the Chern-Pontryagin term.
%(see Fig~\ref{fig:evo_curvatures}).
Figure~\ref{fig:evo_curvatures} shows the evolution of the means, and rms of $^{(3)}R$, $C$ and $P$ with respect $H^2$, across several simulations.
During the decelerated expanding period of pre-inflation, all these quantities decrease in amplitude, but at a smaller rate that $H^2$ and therefore, they experience a modest growth with respect to it. %, up to one order of magnitude above. 
Eventually, in our settings around ${\cal N} \approx 2.5$ efolds, all kinematic energies from the matter and gravity sector fall below the scalar-field potential energy $V(\varphi)$ which becomes dominant. Naturally, this is when cosmic inflation begins. During inflation, the Hubble rate becomes nearly constant, while the Ricci, Weyl and Chern-Pontryagin keep diminishing at an exponential rate of $^{(3)}R \propto a^{-3/2}$, ${C \propto a^{-4}}$, $ {P\propto a^{-5}}$. After 7 or 8 efolds of inflation, at ${\cal N} \approx 10$ in our simulations, $C, ~ P \ll {\cal O}(-10) H^2$, yielding a nearly flat and isotropic Universe which evolution can be accurately described by Friedmann equations of a FLRW Universe. The full duration of inflation can therefore be estimated by the value of the scalar-field at a given inflationary region. In these studied scenarios, inflation starts at $\varphi \approx 18.9~\mpl$, which would yield an inflationary phase of $\mathcal{N}_{\rm inf} >10^3$ efolds. In any case, after to the necessary $\mathcal{N}_{\rm inf} \sim 60$ efolds, all quantities fall below the CMB constraints. 

An alternative way to investigate this process is to study the energy densities contributions, as defined in Eqs.~(\ref{eq:friedman}) and (\ref{eq:densities}).
Figure~\ref{fig:pics_density}, show two-dimensional snapshots at different e-folding times. The top panel shows the relative local variation of the Hubble rate, and the middle and bottom panels show the gravitational curvature and shear, respectively, over the scalar-field energy density. These plots show the rich inhomogeneous dynamics taking place 
in the firsts $\sim 2.5$ efolds,
%between ${\cal N} \sim 0$--$2.5$ efolds,
 and the contrast with the later homogenisation after inflation has started. Remarkably, we observe regions strongly dominated by shear or curvature, and that those regions where the $^{(3)}R$ becomes positive (or $\rho_R$ negative), might start a short period of contraction if gravitational shear is also co-dominant with the scalar-field gradients and kinematics. These contracting regions are those represented in red color in the top panel and they are transient, i.e. they appear and disappear across the simulated grid during the highly inhomogeneous period. In our studies, see also \cite{Joana2020}, when these contracting regions form at sub-Hubble scales, they never dominate the mean expansion rate of that Hubble volume and the high medium pressure prevents their collapse into black holes. After the curvature oscillates, and loses strength, the contracting regions revert into a positively expanding rate, similarly as what it was shown in Ref.~\cite{Clough:2016ymm}.

\subsection{On super-Hubble modes}

The formation of black holes is still possible when super-Hubble scalar field gradients enter the Hubble scales, however they do not prevent inflation to start as previously shown in ~\cite{East:2015ggf,Clough:2016ymm}. On the contrary, they tend to trap oscillating sub-Hubble modes in the collapsing region, and therefore easing the homogenisation in the near region outside the black hole, speeding the transition to inflation. 
We also tested the case with super-Hubble modes in the metric tensors, and as expected,  they remain nearly constant at super-Hubble scales, becoming only ''dynamical" after Hubble entry, naturally sourcing large fluctuations in the extrinsic curvature tensor. In this work, we assume that large kinetic values in the metric, i.e. $\tilde A\ij$, are only physical at around Hubble scales and below. This is important because is the main difference between our studies on initial conditions and those examined in Ref.~\cite{Ijjas:2024oqn}. 
\\
In their work, they investigate initial configurations with strong super-Hubble modes in the extrinsic curvature tensor, analogue to the $\tilde A\ij$ in BSSN variables. This corresponds to setting large kinematic modes of the metric tensor because of $\partial_t \tilde\gamma\ij \propto \tilde A\ij$. Indeed, these initial conditions impose dynamical modes that evolve at super-Hubble scales that induce distortions of spacetime in a non-causal manner. If these modes are strong, they can even induce contraction along a given spatial dimension, overtaking a positive expansion rate from $H \approx - K/3$. In these conditions, scalar field gradients and gravitational scalar curvature might grow out of control, especially because restoration forces induced by gradients in the metric are weak at super-Hubble scales. 
%\\
The disparity between the conclusions of this paper and the ones in \cite{Ijjas:2024oqn} is therefore likely to be related to this issue, however, further investigations are necessary for full clarification. 
These questions on the ''generality" of the initial configurations, as well as the overall shape~\cite{Linde:2004nz,Linde:2017pwt} of the pre-inflationary Universe is beyond the scope of this paper, and they are left for future studies.

\subsection{Can cosmic inflation be ever prevented?}

One possible way that inflation could be prevented is if inhomogeneities could drive the overall expanding regions into contracting ones and therefore make the Universe to re-collapse to the quantum gravity period before any inflation could possibly start.  However, our recent works strongly suggest that inhomogeneities cannot make the Universe collapse at super-Hubble scales, and can only form either transient contracting regions at sub-Hubble scales, or pre-inflationary black holes at Hubble crossing. Despite that inflation will not start inside of formed black holes, the formation of these objects do not prevent cosmic inflation from starting in its surroundings. On the contrary, they tend to facilitate its transition.

Another possibility that could make inflation fail, would be the case when the field is driven ''out" of the inflationary part of the potential before the energy content becomes dominated by it. As shown in ~\cite{Clough:2016ymm,Elley:2024alx}, this is certainly a problem for inflationary models working at sub-Planckian field values, where scalar field velocities and/or gradients might ''drag" the field to the potential minima, and therefore preventing or shortening any period of inflation%
\footnote{
Previous studies have shown that small-field inflationary models are not robust to the initial conditions. A word of caution, this should be viewed such that the onset of a successful inflationary epoch is very sensitive to initial inhomogeneities, and not as a \textit{no-go} for this type of models. Indeed, it has been shown that
regions that should yield inflation in the slow-roll limit, might fail if kination or scalar field gradients are considered. However, the opposite can also be realised, i.e. regions that initially are located outside the slow-roll part of the potential get dragged into the successful inflationary region, especially if strong field velocities are considered~\cite{Chowdhury:2019otk}. Therefore, the initial condition problem for small-field inflationary models cannot be disregarded easily, and needs to be considered in detail for particular cases. 
}. 
However, for models operating at super-Planckian field values this is not a problem. Indeed, because gradients tend to equalise the scalar field towards its mean value, and large field velocities can only displace the field up to $ \Delta\varphi \sim 1~ \mpl$, it is only necessary that the inflationary part of the potential is large enough to overcome these effects. For a $m\varphi^2$ model this would correspond to patches with a mean field of $\varphi_ini \gtrsim 6~\Mpl$, and roughly the same value for Starobinski-like models, in order to have an inflationary period of  ${\cal N}_{\rm inf} \gtrsim 60$ efolds.

Last, another potential danger for inflation to begin, would be the decay of the inflaton field into other fields. This is a very specific problem, which depends on many theoretical ansatzs. In this work, we ignored any potential coupling with additional fields, but in Ref.~\cite{Joana2022} the effects of auxiliary fields were investigated for the particular case of non-minimal Higgs inflation. In that situation, the non-minimal coupling of the inflaton protects from the dynamics of the auxiliary field becoming energetically dominant. However, even for simple models, where auxiliary fields fluctuations could dominate the energy content, the overall expansion of the Universe should make those perturbations to decay, and eventually the scalar-field(s) potential should become dominant. As long as this potential is flat enough to allow inflation at super-Planckian field values, inflation should begin in a similar manner as shown in this work.

\subsection{Observables from the pre-inflation phase?}

\begin{figure}[t!]
     \includegraphics[width=7.5cm]{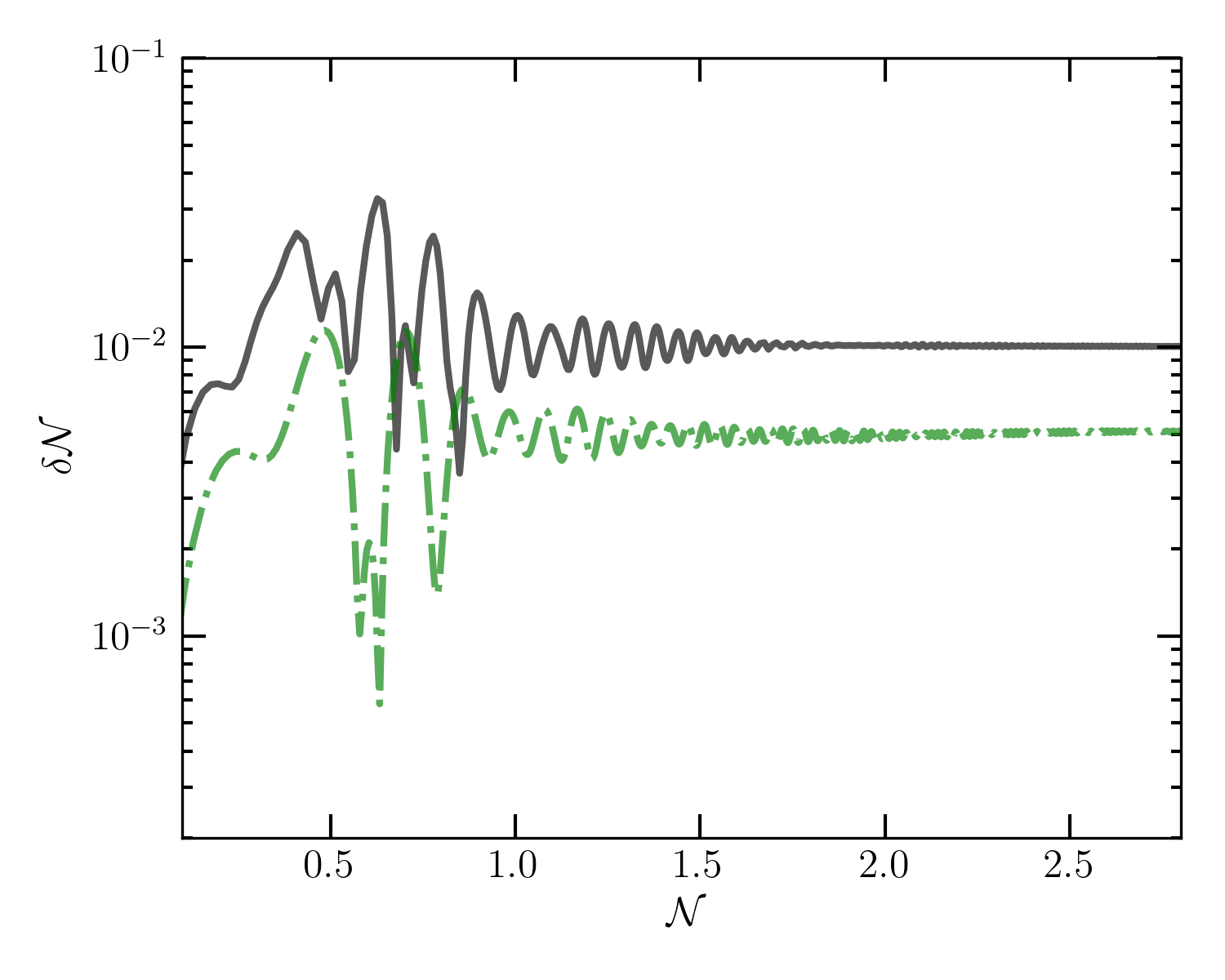} 
     \caption{Evolution of the mean scalar gravitational curvature, which amplitudes freeze at super-Hubble scales after inflation onset, around ${\cal N} \approx 2.5$. 
 	Same colour code as in Fig.~\ref{fig:evo_curvatures}.    
     }
     \label{fig:evo_zeta}
 \end{figure}

To finalise, we also investigate the evolution of gravitational curvature perturbations, shown in Figure~\ref{fig:evo_zeta}. The figure confirms that during the inhomogeneous period of pre-inflation curvature perturbations are generated, and eventually they freeze-out during inflation when they are stretched at super-Hubble scales. In principle, this opens a possible (but unlikely) observational window to measure these perturbations if they re-enter at a late time in our Universe. Evidently, that would only be possible if the duration of inflation is very much tuned to just  explain the observed homogeneity of the CMB, i.e. $\sim 60 $ efolds. However possible, this situation would require of certain fine-tuning in the models, but could potentially be seen as features in the large-scale structure and explain some of the observed anomalies \cite{Secrest:2022uvx,10.1093/mnrasl/slae039} .

\section{Discussion} \label{sec:Discussion}

This paper is a follow up of previous numerical relativity studies on the topic of initial conditions of inflation. In here, I have investigated the robustness of large-field inflation models to inhomogeneous initial conditions beyond the assumption of conformal flatness, where initial configurations with large fluctuations in the metric tensor have been considered. These results and conclusions should be valid for any inflation model that operates at large field values, including the most favoured models by Planck such as Starobinski inflation, Non-minimally coupled Higgs inflation, etc. 
\ \\

An important remark is that, intrinsically to the studied initial configurations, I have assumed certain characteristics for the pre-inflationary Universe that might condition the results. These assumptions are the following:
\begin{itemize}
\item[i)] the pre-inflationary Universe is expanding on average, and the matter content is dominated by a single scalar field,  %\\
\item [ii)]the Universe is governed by Einstein's field equations of General Relativity, and can be described classically, where thermal fluctuations dominate and therefore quantum effects can be ignored,  %\\
\item[iii)] the overall energy scales are below Planckian energies, so no quantum theory of gravity is necessary, % \\
\item[iv)] the pre-inflationary Universe contains many Hubble volumes, and super-Hubble fluctuations are expected to exist. 
\item[v)] dynamics evolve locally, i.e. super-Hubble modes 
remain frozen until nearly Hubble entry. 
\end{itemize}
%\ \\

This and previous works have consistently shown that large inhomogeneities, either in the scalar or gravitational field sector, do not prevent cosmic inflation from occurring. We show that all types of gradients eventually fall down the energy scales at which inflation takes places and therefore cosmic inflation begins without need of fine-tuning. 
From these simulations, I conclude that for large-field inflation models, beginning  inflation is an unavoidable outcome as long as in the primordial Universe there existed at least one Hubble volume ${\cal V}_H$ 
where the following characteristics are met:

\begin{itemize} 
\item the initial ${\cal V}_H$ is in average expanding, i.e. ${\left. 
\langle H\rangle\right._{\small {\cal V}_H} > 0}$.
\item the mean value of the scalar-field in $\cal V_H$ is within the slow-roll region of $V(\varphi)$, allowing for at least $60$ efolds in the homogeneous limit. 
\end{itemize}

When the previous assumptions are satisfied, cosmic inflation safely begins, and consequently homogenise, anisotropise and flatten the Universe well beyond super-Hubble scales. We particularly tested that large metric fluctuations, characterized by large Weyl curvatures, are also flattened within a few tens of efolds, and therefore the inflationary mechanism remains a successful cosmic smoother of inhomogeneous pre-inflationary inhomogeneities, including initial settings consistent within the Weyl Curvature Hypothesis. In conclusion, cosmic inflation successfully homogenizes the primordial Universe and therefore can successfully explain the large-scale homogeneity and flatness of today's Universe.
\\ 

To finalize, I acknowledge that the presence of non-causal kinematics at super-Hubble scales, in violation of our fifth assumption described above, can clearly be \textit{dangerous} for starting inflation, as the investigations in~\cite{Ijjas:2024oqn} interestingly shows. However, in my opinion, these scenarios should be considered as specific conditions that might spoil the onset of inflation, rather than being part of the norm of generic ones. Despite that we do not know what kind of conditions might originate from an early quantum gravity phase, and taking into account our limitations on studying the initial condition problem of inflation using numerical general relativity, at the classical level, it should arguably be assumed that in generic initial configurations causality and locality should be preserved.  
\\

\section{Acknowledgments}

The author warmly thanks Christophe Ringeval, Sebastien Clesse and Shi Pi for helpful discussions. I would also like to thank the GRChombo team (http://www.grchombo.org/) for their work on the code. This work is supported by the National Key Research and Development Program of China Grant No. 2021YFC2203004.  
C.J. is also funded by the NSFC grant Num. 12347132, he also wishes to  thank for the hospitality received while visiting the ULB and UCLouvain, and acknowledges the usage of CURL Cosmo clusters at the University of Louvain, funded by the “Fonds de la Recherche Scientiﬁque - FNRS” under Grant Num. T.0198.19. 
Analysis and visualizations employed the Visit \cite{10.5555/2422936} and yt-project \cite{Turk_2010} software packages.
%

%
%% Paciència, perseverança, perspectiva. C.P.C
%
%% When a real and final catastrophe should befall us in Palestine the first responsible for it would be the British and the second responsible for it would be the Terrorist Organizations build up from our own ranks. I am not willing to see anybody associated with those mislead and criminal people. A.E
%

\appendix

\section{BSSN formalism of numerical relativity}
In this work, we solve the BSSN formulation of the Einstein equations using \texttt{GRChombo}~\cite{Clough_2015,Andrade:2021rbd}, a multipurpose numerical relativity code. In the context of the 3+1 decomposition of General Relativity, the line element reads
\be\label{timeline_AP}
\rr d s^2 = - \alpha^2 \rr d t^2 + \gamma_{ij}(\rr d x^i + \beta^i \rr d t)(\rr d x^j + \beta^j \rr d t)
\ee
where $\gamma\ij$ is the metric of the 3-dimensional hypersurface, and the lapse and shift gauge parameters are given by $\alpha(t)$ and $\beta^i(t)$ {respectively}. A further conformal decomposition of the 3-metric follows,
\be
\gm\ij = \frac1\chi \tgm\ij = \psi^4\tgm\ij \quad \text{with } \text{ det}(\tgm\ij) = 1 ~, 
\ee
where $\chi$ and $\psi$ are two different parametrisations of the metric conformal factor. While the former is used during the temporal integration, the latter is preferred when constructing the initial conditions. The extrinsic curvature is thus split in $\tA\ij$ and $K$, respectively, the conformal traceless part and its trace, 
\be
K\ij = \frac1\chi \left( \tA\ij +\frac13\tgm\ij K\right)~.
\ee
In addition, the first spatial derivatives of the metric are considered as dynamical variables
\be
\tilde\Gamma^i \equiv \tgm^{jk} \tilde\Gamma^i_{jk} = - \partial_j\tgm\ij ~,
\ee
where $ \tilde\Gamma^i_{jk} $ are the Christoffel symbols associated with the conformal metric $ \tilde\gamma_{ij} $.
\ \\

The evolution equations for the BSSN variables are then given by 
\begin{align} 
&\partial_t\chi=\frac{2}{3}\,\alpha\,\chi\, K - \frac{2}{3}\,\chi \,\partial_k \beta^k + \beta^k\,\partial_k \chi ~ , \label{eqn:dtchi2} \\
&\partial_t\tilde\gamma_{ij} =-2\,\alpha\, \tA_{ij}+\tilde \gamma_{ik}\,\partial_j\beta^k+\tilde \gamma_{jk}\,\partial_i\beta^k \nonumber \\
&\hspace{1.3cm} -\frac{2}{3}\,\tilde \gamma_{ij}\,\partial_k\beta^k +\beta^k\,\partial_k \tilde \gamma_{ij} ~ , \label{eqn:dttgamma2} \\
&\partial_t K = -\gamma^{ij}D_i D_j \alpha + \alpha\left(\tilde{A}_{ij} \tilde{A}^{ij} + \frac{1}{3} K^2 \right) \nonumber \\
&\hspace{1.3cm} + \beta^i\partial_iK + 4\pi\,\alpha(\rho_{\varphi}+ S) \label{eqn:dtK2} ~ , \\ 
&\partial_t\tA_{ij} = \left[- \chi D_iD_j \alpha + \chi \alpha\left( R_{ij} - 8\pi\, \,S_{ij}\right)\right]^\textrm{TF} \nonumber \\
&\hspace{1.3cm} + \alpha (K \tA_{ij} - 2 \tA_{il}\,\tA^l{}_j) \nonumber \\
&\hspace{1.3cm} + \tA_{ik}\, \partial_j\beta^k + \tA_{jk}\,\partial_i\beta^k \nonumber \\
&\hspace{1.3cm} -\frac{2}{3}\,\tA_{ij}\,\partial_k\beta^k+\beta^k\,\partial_k \tA_{ij}\, \label{eqn:dtAij2} ~, \\ 
&\partial_t \tilde \Gamma^i=2\,\alpha\left(\tilde\Gamma^i_{jk}\,\tA^{jk}-\frac{2}{3}\,\tilde\gamma^{ij}\partial_j K - \frac{3}{2}\,\tA^{ij}\frac{\partial_j \chi}{\chi}\right) \nonumber \\
&\hspace{1.3cm} -2\,\tA^{ij}\,\partial_j \alpha +\beta^k\partial_k \tilde\Gamma^{i} \nonumber \\
&\hspace{1.3cm} +\tilde\gamma^{jk}\partial_j\partial_k \beta^i +\frac{1}{3}\,\tilde\gamma^{ij}\partial_j \partial_k\beta^k \nonumber \\
&\hspace{1.3cm} + \frac{2}{3}\,\tilde\Gamma^i\,\partial_k \beta^k -\tilde\Gamma^k\partial_k \beta^i - 16\pi\,\alpha\,\tilde\gamma^{ij}\,S_j ~ , \label{eqn:dtgamma2}
\end{align} 
%%%
where the superscript $\rm{TF}$ denotes the trace-free parts of tensors, with $R\ij$ being the (3-dimensional) Ricci tensor.  The 3+1 decomposition of the energy-momentum tensor $T^{\mu\nu}$ gives
\bea \label{3+1sources_AP}
 \rho_\varphi &=& n^\mu n^\nu T_{\mu\nu} ~,\\
 S_i &=& -\gamma^{\mu}_i n^\nu T_{\mu\nu} ~,\\ 
 S_{ij} &=& \gamma^{\mu}_i \gamma^{\nu}_j T_{\mu\nu} ~,\\ 
 S &=& \gamma\IJ S\ij ~,
\eea
where $n^\mu=(1/\alpha, -\beta^i/\alpha)$ is the unit normal vector to the three-dimensional slices.

The Hamiltonian and Momentum constraints, 
\begin{align}
\mathcal{H} & = R + K^2-K_{ij}K^{ij}-16\pi \rho= 0\, , \label{eqn:HamSimp} \\
\mathcal{M}_i & = D^j (K_{ij} - \gamma_{ij} K) - 8\pi S_i =0\, , \label{eqn:MomSimp}
\end{align}
where $R$ is the Ricci scalar, are only solved explicitly when constructing initial data. However, they are also monitored during the time evolution in order to ensure that there is no significant deviations from General Relativity. 
\ \\
%\subsection{Gauge choice and singularity avoidance }%

The gauge parameters are initially set to $\alpha=1$ and $\beta^i=0$ and then evolved in accordance with the \textit{moving puncture gauge} \cite{Baker_2006, Campanelli_2006}, for which evolution equations are
\begin{eqnarray}
\partial_t \alpha &=& -\eta_\alpha \alpha \left( K - \langle K \rangle \right) + \,\beta^i\partial_i \alpha \ , \label{eqn:alphadriver}\\
\partial_t \beta^i &=& B^i\, \label{eqn:betadriver},\\
\partial_t B^i &=& \frac34\, \partial_t \tilde\Gamma^i - \eta_B\, B^i\ \,, \label{eqn:gammadriver}
\end{eqnarray}
where the constants $\eta_\alpha$ and $\eta_B$ are conveniently chosen to improve the numerical stability. This way, $\alpha$ and $\beta^i$ are boosted in the problematic regions with strongly growing extrinsic curvature and spatial derivatives of the three-metric $\tilde \gamma_{ij}$. 
The goal of this gauge is to prevent the code from resolving the central singularity of any black hole that may eventually form, as well as to prevent coordinate singularities on converging geodesics.

\subsection{Scalar field equations}
For the Einstein frame canonical scalar field $\varphi$, the energy-momentum tensor is given by
\begin{equation}
T_{\mu\nu} = \left( \partial_\mu \varphi\, \partial_\nu \varphi - \frac{1}{2} g_{\mu\nu}\, \partial_\lambda \varphi \, \partial^\lambda \varphi\right) - g_{\mu\nu} V(\varphi) \,
\end{equation}

The scalar field dynamics is governed by the the Klein-Gordon equation, split into two first order equations for the field and its momentum $\Pi_{\rm M}^I$
\begin{align}
\partial_t \varphi &= \alpha \Pi_{\rm M} +\beta^i\partial_i \varphi \label{eq:dtvarphi} ~ , \\
\partial_t \Pi_{\rm M} &= \beta^i\partial_i \Pi_{\rm M} + \alpha\partial_i\partial^i \varphi + \partial_i \varphi \, \partial^i \alpha \\
& \ +\alpha \left( K\Pi_{\rm M}-\gamma^{ij}\Gamma^k_{ij}\partial_k \varphi - \frac{d}{d\varphi}V(\varphi) \right) ~ ,
\end{align}

\section{Construction of initial data}

The initial conditions for the metric tensor, that has been used in our simulations, have been constructed by providing the parameters listed in Table~\ref{tab:initial_data}.

\begin{table}[hb!]
Initial data parameters
\begin{tabular}{ll} 
% sim08
\\ \hline \hline
%mix (I):  $\quad$ 
mix:  $\quad$  
	   &  $a_0  = b_0 = c_0 = 3\cdot 10^{-3}$,  $a_1 = b_1  = c_1 = 10^{-3}$,   \\
(Fig.~\ref{fig:pics_density})    
       &  $a_2 = b_2 = c_2 = 1$,       $a_3 = b_3 = c_3 = \frac13 \pi$,  \\
       &  $d_1 = 0.1$, $f_1=0.01$, $g_1 = 0.033$,      \\
       &  $d_2 = f_2 =g_2= 1$, $d_3 = \frac13 \pi$,   $f_3 =  \frac 12 \pi$,   $g_3 = \pi \frac 9{10}$ \\[2mm]
\hline \hline 
off-diagonal
	   &  $a_0  = b_0 = c_0 = a_1 = b_1  = c_1 = 0$,   
	   \\
 (I): 	  
       &  $a_2 = b_2 = c_2 = 1$,       $a_3 = b_3 = c_3 = \frac13 \pi$,  \\
       &  $d_1 = 0.1$, $f_1=0.001$, $g_1 = 0.03$,      \\
       &  $d_2 = f_2 =g_2= 1$, $d_3 = \frac13 \pi$,   $f_3 =  \frac 12 \pi$,    $g_3 = \pi \frac 9{10}$ \\[2mm]
\hline \hline 
% sim09        
diagonal (I):  
	   &  $a_0  = b_0 = c_0 = 0.06$,  $a_1 = b_1  = c_1 = 0.02$,   
	   \\
       &  $a_2 = b_2 = c_2 = 1$,       $a_3 = b_3 = c_3 = \frac13 \pi$,  \\
       &  $d_1 = f_1= g_1 = 0 $,      \\
       &  $d_2 = f_2 =g_2= 1$, $d_3 = \frac13 \pi$,   $f_3 =  \frac 12 \pi$,    $g_3 = \pi \frac 9{10}$ \\[2mm]
\hline \hline 
%
% sim06b       
off-diagonal  
	   &  $a_0  = b_0 = c_0 = a_1 = b_1  = c_1 = 0$,   
	   \\
 (II):
       &  $a_2 = b_2 = c_2 = 1$,       $a_3 = b_3 = c_3 = \frac13 \pi$,  \\
       &  $d_1 = 0.01$, $f_1=10^{-4}$, $g_1 = 3\cdot10^{-3} $,      \\
       &  $d_2 = f_2 =g_2= 1$, $d_3 = \frac13 \pi$,   $f_3 =  \frac 12 \pi$,    $g_3 = \pi \frac 9{10}$ \\[2mm]
\hline \hline 
%
% sim07        
diagonal (II): 
	   &  $a_0  = b_0 = c_0 = 0.01$,  $a_1 = b_1  = c_1 = 0.01$,   
	   \\
       &  $a_2 = b_2 = c_2 = 1$,       $a_3 = b_3 = c_3 = \frac13 \pi$,  \\
       &  $d_1 = f_1= g_1 = 0 $,      \\
       &  $d_2 = f_2 =g_2= 1$, $d_3 = \frac13 \pi$,   $f_3 =  \frac 12 \pi$,    $g_3 = \pi \frac 9{10}$ \\[2mm]
\hline \hline 
\end{tabular}
\caption{ Initial parameter sets used to compute the initial data of the simulations used throughout this work.  They correspond to the colour-code lines used in the figures: black, red, blue, green, purple, respectively from top to bottom.
\label{tab:initial_data}
}
\end{table}

\begin{figure*}[b!]
    \centering 
    \includegraphics[width=15.5cm]{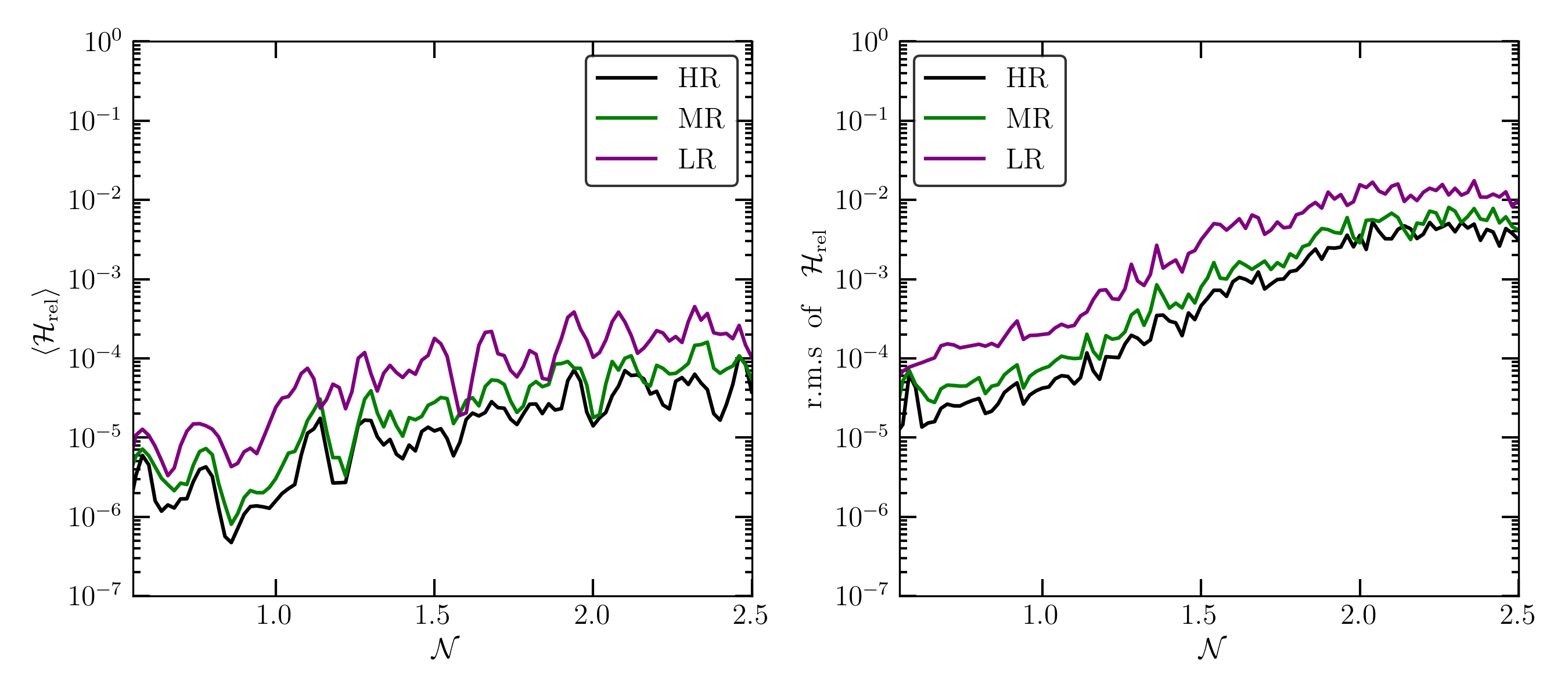} 
    \caption{Convergence testing. Shows the mean and r.m.s. of the relative Hamiltonian constraint using different grid resolutions: $\mathrm{HR} = (280)^3$,  $\mathrm{MR} = (220)^3$, $\mathrm{LR} = (160)^3$. Same initial conditions as shown in Fig. ~\ref{fig:pics_density}.    
    }
    \label{fig:convergence}
\end{figure*}

%\newpage

\section{Code validation and convergence tests}
The validation of the code is done by monitoring the constraint equations.  The relative Hamiltonian constraint is defined by
%$
\be \label{eq:HamRel}
\mathcal{H}_{\rm REL}^2  \equiv \frac
{ \left(
 {\, ^{(3)}R}  -  \tilde A\IJ \tilde A\ij 
+ \frac23 K^2   -  16\pi \rho_{\varphi} \right)^2}
{  {\, ^{(3)} R} ^2 +  \lp \tilde A\IJ \tilde A\ij \rp ^2 
+ \lp  \frac23 K^2 \rp ^2  + \lp 16\pi \rho_{\varphi}\rp ^2 }
\ee
where the numerator in Eq.(\ref{eq:HamRel}) is just the Hamiltonian constraint expressed in BSSN variables. 
The $\mathcal{H}_{\rm REL}$ quantity has been computed for all simulations and monitored that their values have remain low through the simulation time. Convergence tests using different grid-size resolutions are shown in Fig.~\ref{fig:convergence}.

\newpage
\clearpage

\bibliographystyle{apsrev4-1}
\bibliography{biblio.bib}

%\newpage
%\clearpage

\end{document}